\documentclass[twocolumn,prb, showpacs]{revtex4}
\usepackage{graphicx}
\usepackage{dcolumn}
\usepackage{bm}
\begin{document}
\title{Searching for Stable Na-ordered Phases in Single Crystal Samples of $\gamma$-Na$_{x}$CoO$_{2}$}
\author{G. J. Shu$^1$, Andrea Prodi$^2$, S. Y. Chu$^3$, Y. S. Lee$^2$, H. S. Sheu$^4$ and F. C. Chou$^{1,4,\dagger}$}
\affiliation{$^1$Center for Condensed Matter Sciences, National Taiwan University, Taipei 10617, Taiwan,\\
$^2$Department of Physics, Massachusetts Institute of Technology,Cambridge, MA 02139\\
$^3$Center for Materials Science and Engineering, Massachusetts Institute of Technology,Cambridge, MA 02139\\
$^4$National Synchrotron Radiation Research Center, HsinChu 30076,
Taiwan\\}
\date{\today}
\begin{abstract}
We report on the preparation and characterization of single crystal
$\gamma$ phase Na$_{x}$CoO$_{2}$ with $0.25 \leq x \leq 0.84$ using
a non-aqueous electrochemical chronoamperemetry technique.  By
carefully mapping the overpotential versus $x$ (for $x < 0.84$), we
find six distinct stable phases with Na levels corresponding to
$x \sim$ 0.75, 0.71, 0.50, 0.43, 0.33 and 0.25.  The composition
with $x\simeq0.55$ appears to have a critical Na concentration which
separates samples with different magnetic behavior as well as
different Na ion diffusion mechanisms. Chemical analysis of an aged
crystal reveals different Na ion diffusion mechanisms above and
below $x_c \sim 0.53$, where the diffusion process above $x_c$ has a
diffusion coefficient about five times larger than that below $x_c$.
The series of crystals were studied with X-ray diffraction,
susceptibility, and transport measurements. The crystal with $x = 0.5$
shows a weak ferromagnetic transition below $T=27$ K in addition to
the usual transitions at $T = 51~K$ and $88~K$. The resistivity of the
Curie-Weiss metallic Na$_{0.71}$CoO$_{2}$ composition has a very low
residual resistivity, which attests to the high homogeneity of the
crystals prepared by this improved electrochemical method. Our
results on the various stable crystal compositions point to the
importance of Na ion ordering across the phase diagram.
\end{abstract}
\pacs{61.10.Nz, 61.50.Nw, 66.30.-h, 75.20.Hr, 75.30.Cr, 71.27.+a}
\maketitle
\section{\label{sec:level1}Introduction\protect\\ }
The lamellar cobaltates Na$_{x}$CoO$_{2}$ have attracted much
attention since the discovery of superconductivity in the hydrated
composition Na$_{0.3}$CoO$_{2}$$\cdot$1.4H$_2$O.\cite{Takada2003}
The complete phase diagram of Na$_{x}$CoO$_{2}$ shows a variety of
rich phenomena, such as A-type antiferromagnetic (AF) ordering for
$x \sim$ 0.75, large thermopower enhancement for $x = 0.85$, and AF
ordering with a metal-to-insulator transition for
$x=0.5$.\cite{Bayrakci2005, Foo2004, Bobroff2006,
Gasparovic2006, Lee2006} Na$_x$CoO$_2$ is composed of alternating
layers of Na and CoO$_2$, where the Na ions are surrounded by six
oxygens which form a prismatic cage in the $\gamma$
phase.\cite{Delmas1981} Sodium de-intercalation in powder samples
has been effectively performed via chemical extraction, such as a
topochemical process using a Br$_2$/acetonitrile
solution.\cite{Takada2003} We have demonstrated previously an
electrochemical method as an alternative route to produce
homogeneous compounds using a controlled overpotential (above the
open circuit potential).\cite{Chou2005} Using an aqueous
NaOH/H$_2$O solution as the electrolyte, high quality single
crystals were obtained, and measurements of their conducting,
superconducting and magnetic properties have been
reported.\cite{Balicas2005, Chou2004} However, detailed studies of
the voltage versus Na concentration diagram for $x < 1/2$ have been
hindered due to difficulties related to an oxygen evolution
side-reaction in the aqueous solution.\cite{Chou2005}

Na ordering - in particular its impact on the geometry of the Fermi
surface - is a key issue that requires further study in order to
better understand the transport and magnetic
properties.\cite{Balicas2005, Singh2000, Bobroff2006, Lee2005a, Zhou2007} Local density
approximation (LDA) calculations indicate that a band with  a$_{1g}$
character should create a large hexagonal Fermi surface centered
around the $\Gamma$ point.\cite{Singh2000}  The existence of such a
large Fermi surface has been verified from angle-resolved
photo-emission(ARPES).\cite{Yang2004,Hasan2004}  However, subtle
effects due to possible superstructure formation and/or strong
electron correlation remain elusive, such as the missing small hole
pockets predicted by LDA calculation.  Bobroff {\em et al.} proposed
a nesting scenario based on an orthorhombic reduced Brillouin zone
caused by Na ordering to interpret the antiferromagnetic and
metal-to-insulator transitions found in
Na$_{0.5}$CoO$_2$.\cite{Bobroff2006}  A reconstructed Fermi surface
may also explain the small hole pocket found in Na$_{0.5}$CoO$_2$
and Na$_{0.3}$CoO$_2$ by Shubnikov-de Haas oscillation
measurements.\cite{Balicas2005, Balicas2006}  Several superstructures have been observed and modeled from electron diffraction studies of
Na$_x$CoO$_2$ with $x$ between 0.15 and 0.75.\cite{Zandbergen2004}
Calculations based on electrostatic interaction between Na ions
alone have been used to propose several types of Na ordering
patterns.\cite{Zhang2005} Other calculations point to the importance
of the coupling between Na-vacancy ordering and Co$^{3+}$/Co$^{4+}$
charge ordering, indicating that Na$^{+}$--Na$^{+}$ repulsion and
the Na(1)--Na(2) site energy difference alone is not enough to
describe the observed superlattices.\cite{Meng2005} A recent neutron
diffraction study and associated numerical simulations have
elucidated aspects of the Na ordering in single crystals of
Na$_x$CoO2 with x $\sim$ 0.75 - 0.92.\cite{Roger2007a}

Much still remains to be understood regarding the possible Na
ordering patterns across the phase diagram. Currently, Na ordering
has been studied mostly in samples with $x = 1/2$ and $x\simeq 3/4$.
However, reproducibly preparing precise stoichiometries of these
phases is not straightforward. Many early studies of samples with $x
= 0.7 - 0.9$ did not pay much attention to the exact ordering of Na
ions and often assumed substantial defect formation due to Na loss
at high temperatures.\cite{Motohashi2001}  Although
Na$_{0.5}$CoO$_2$ shows the highest ordering and phase stability
among the Na$_x$CoO$_2$ series, phase inhomogeneity and phase
separation have been discussed based on electron microscopy studies
of Na$_{0.5}$CoO$_2$.\cite{Zandbergen2004, Yang2004a} A reliable
procedure for controlling the Na content and the rate of
de-intercalation of single crystal samples of Na$_x$CoO$_2$ is
highly desirable, especially in light of the possibility of
producing Na ordered phases.

We find that an electrochemical method which carefully controls the
overpotential at the sample surface is a particularly reliable route
that can approach the real equilibrium state and, moreover, is
suitable for single crystal samples. The topochemical method is an
effective method to produce powder samples as reported
previously.\cite{Schaak2003}  However, the Na level in powders is
more susceptible to Na loss due to the much shorter diffusion path
and significantly larger surface-to-bulk ratio.   In fact, we find
that over time, Na diffusion under normal storage conditions can
affect surface layer stoichiometry as deep as 50 $\mu$m as we
discuss below.  Here, we have used electrochemistry in a non-aqueous
solution to carefully prepare a series of homogeneous single
crystals.  Characterization of these samples reveals that six
distinctively stable phases fall in the stoichiometry of $x \sim$
0.75, 0.71, 0.50, 0.43, 0.33 and 0.25.  Most phases are clearly near
simple commensurate filling fractions, while 0.71 and 0.43 require
further modeling and simulation.  Magnetic susceptibility and
transport measurements have been performed on these samples and the
results are discussed.
\section{\label{sec:level1}Experimental\protect\\}
All single crystals in this study were cleaved from a crystal grown
using the floating-zone method as described
previously.\cite{Chou2004, Prabhakaran2004} In order to minimize any
possible Na content differences between the surface and bulk, we
performed all electrochemical experiments using thin crystals with
thicknesses as small as 100 $\mu$m.  The electrochemical cell was
assembled using crystal sample as working electrode, Pt as counter
electrode, and Ag/AgCl with sieved liquid junction as reference
electrode. In contrast to our previous work, a non-aqueous
electrolyte was used: 1N NaClO$_4$ in propylene carbonate. Single
crystals of Na$_{0.84}$CoO$_2$ with size near 2$\times$2$\times$0.1
mm$^3$ were sandwiched into a spring-loaded  cell to make the
working electrode.  A DC voltage $V_{app}$ was applied between the
working and reference electrodes at each applied voltage step,
spanning the range from -0.25 V to 2.5 V relative to the Ag/AgCl
reference. Equilibration is achieved by waiting until the induced
anodic current reaches saturation and maintained for at least 12
hours.
For chemical analysis of the stoichiometry, electron probe
microanalysis (EPMA) was performed on freshly cleaved inner
surfaces. To avoid electron-beam damage, EPMA data were collected
using a low beam energy of 15 keV for 1 min duration on each fresh
spot. Structural and bulk magnetic data were obtained from X-ray
diffraction and SQUID magnetometry measurements, respectively. For
the X-ray diffraction, single crystal samples were oriented, and the
(00L) reflections at high angles were used to determine the c-axis.
In-plane resistivity data of our various samples were obtained by
performing four terminal DC measurements on single crystal samples.
The crystals were cut to $1 mm \times 0.5 mm \times 0.1 mm$ average
size. The leads were attached with DuPont 4922N silver paint on a
cleaved $ab$ surface and dried in air, yielding contact resistances
between $5 - 20 \Omega$.
\section{\label{sec:level1}Results and Analysis\protect\\ }
\begin{figure}
\includegraphics [width=3.3in]{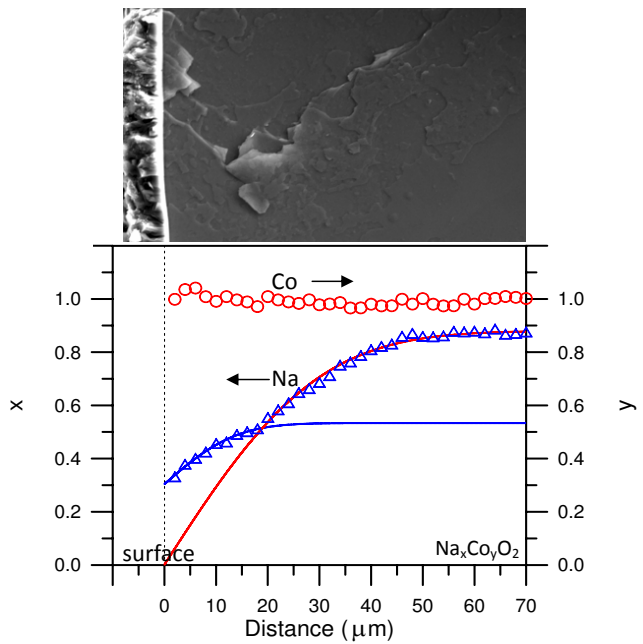}
\caption{\label{fig:diffusion} (Color online)(a) SEM image of an ``aged'' ($\sim$
2.5 years) crystal of Na$_{0.88}$CoO$_2$ cleaved with the $c$-axis
perpendicular to the plane  The left edge corresponds to the surface
of the as-grown rod. (b) EPMA analyzed chemical concentration $x$ as
a function of distance from the crystal surface towards the interior
of the rod. Solid lines are fits with a 1D diffusion model as
described in the text.} \vspace{-5mm}
\end{figure}
\subsubsection{\label{sec:level2}Sodium ion self diffusion}
When the as-grown single crystal of Na$_x$CoO$_2$ with $x > 0.7$ is
not stored properly within a moisture-free environment, a white
precipitate builds up on the surface. We believe the surface Na ions
readily react with the water in the air to form white NaOH
precipitate.  The vacated sites would induce further Na ion self
diffusion. In order to better understand the diffusion mechanism of
Na ions, we have examined the Na distribution profile of an "aged"
single crystal of Na$_{0.88}$CoO$_2$ that has been stored in ambient
environment of average humidity of 55\% for about two and a half
years. The original single crystal was prepared using floating-zone
method as reported previously.\cite{Chou2004}  The "aged" crystal
rod had a layer of white precipitate built up on the surface of the
boule.  We performed EPMA to examine the Na/Co ion distribution on a
cleaved surface ($ab$-plane) which extends radially from the core
area to the edge of the rod as shown in Fig. \ref{fig:diffusion}(b).
An SEM photograph is shown in Fig. \ref{fig:diffusion}(a), where the
crystal surface is defined as the zero position. We find that the Na
concentration rapidly drops from $x \simeq 0.88$ in the interior of
the rod to $x \simeq 0.3$ at the surface layer within a depth of
$\sim$50 $\mu$m. The white layer outside the crystal consists of a
rich concentration of Na (Na/Co $>$ 1.67) and has been confirmed
with X-ray diffraction to be NaOH.
The Na$^+$ ions, with ionic radius $\sim 1.02$ {\AA}, may diffuse
through the rectangular pores of threshold radius $\sim 0.91$ {\AA}
formed by the oxygen ions of the NaO$_6$ prismatic
cage.\cite{Shin2002} Since Na$^+$ ions diffuse along the $ab$-plane
toward the surface boundary through a vacancy hopping mechanism, the
observed Na distribution shown in Fig. \ref{fig:diffusion} would
follow Fick's second law. The diffusion topography $c(r,t)$ and the
diffusion coefficient $D_\circ$ can be described by the diffusion
equation $\frac{\partial c(r,t)}{\partial t}=D_\circ \frac{\partial
^2 c(r,t)}{\partial r^2}$ in one dimension.  Although it is tempting
to fit the whole data set with a single diffusion coefficient, we
note the $c(r,t)$ profile has a pronounced crossover near $x \sim
0.5$.  The best fit is obtained under the assumption of two
different diffusion coefficients above and below the crossover point,
as indicated by the solid lines in Fig. \ref{fig:diffusion}.  Leaving
the boundary conditions of $c(r,t)$ at the crossover point and the
surface edge as free fit parameters, the best fit values of
$D_\circ$ are $\sim$ 3.49 and 0.76 $\times$ $10^{-14}~cm^2/s$ for
diffusion above and below the crossover point, respectively. We note
that this is close to the lower limit of $D_{Li}$ for Li$_x$CoO$_2$
reported in the literature.\cite{Jang2001}  The diffusion
coefficients extrapolated from single crystals are less susceptible
to errors due to geometric factors and electrolyte permeation,
compared to powder samples.

Interestingly, the observed crossover point is $x_c \sim 0.53$, which
is just above half filling. The end point at the surface boundary is
near $x \sim 0.30$, right below 1/3 filling. The fit value of
$D_\circ$ for $x > 0.53$ is nearly five times of that from below.
This indicates two distinctly different types of diffusion species
dominated above and below half filling, which are most likely to be
Na vacancy and Na$^+$ ion, respectively. For Na$_x$CoO$_2$ with x
$>$ 0.5, vacancy ordering is expected as demonstrated by the
di-vacancy or tri-vacancy ordering patterns proposed by Roger {\em
et al.}\cite{Roger2007a}  In the multi-vacancy model, energy is
gained through Na(2) vacancy cluster formation plus promotion of
additional sodium ions from the preferred Na(2) site to Na(1) site.
The larger diffusion coefficient for vacancy-dominated diffusion is
also reflected in the lower overpotential required in
electrochemical de-intercalation process, as will be discussed below.
The concentration near the surface boundary is very close to $x =
1/3$, which is exactly the Na concentration that becomes
superconducting after fully hydrated.\cite{Takada2004}  The energy
gap between $x = 1/3$ and $x = 1/4$ appears to be large enough to
prevent additional Na loss from the surface. The crossover point
near $x_c = 0.53$ also separates the two main classes of magnetic
behavior within Na$_x$CoO$_2$, i.e. the Curie-Weiss metal above and
Pauli paramagnetic metal below.\cite{Foo2004}  The observed
different diffusion mechanisms and thus different Na (vacancy)
ordering patterns have shown profound influence on the magnetic and
conducting properties of Na$_x$CoO$_2$.
\begin{figure}
\includegraphics[width=3.3in]{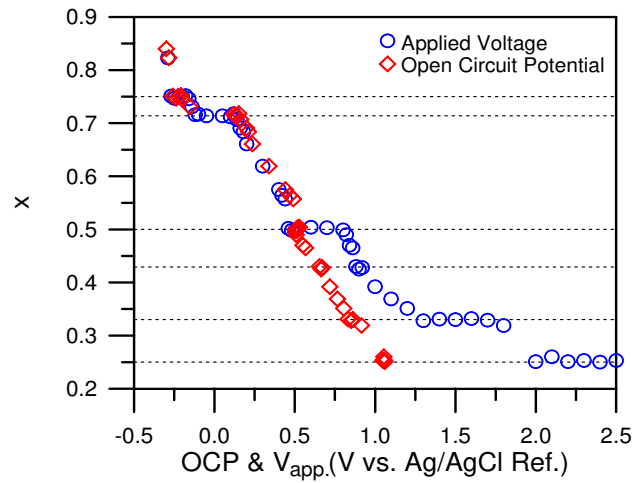}
\caption{\label{fig:deintercalation} (Color online) Na concentration $x$ versus
applied voltage $V_{app}$ and open circuit potential (OCP), where
$x$ is obtained from EPMA analysis.} \vspace{-5mm}
\end{figure}
\subsubsection{\label{sec:level2}Electrochemical de-intercalation}
The electrochemical de-intercalation method relies on the sharp
concentration gradient at the sample/electrolyte interface.  Na ions
diffuse to the surface layer due to the non-zero overpotential until
a new equilibrated surface chemical potential is reached.  A plot of
of the Na concentration (determined using EPMA) versus the applied
electrochemical voltage ($V_{app}$) is shown in
Fig. \ref{fig:deintercalation} for our single crystal samples. As the
applied voltage increases, the Na content decreases similar to
previously reported results on powder samples.\cite{Chou2005,
Delmas1981}  The stability of the resulting phases is evidenced by
the wide plateau near regions of $dV_{app}/dx$ = 0 at $x$ near 0.71,
0.5, 0.33 and 0.25.

Crystals prepared with applied voltages between 0.5 and 0.8 V show
nearly constant Na content of $x = 0.5$. In addition, higher
voltages near 1.3 -- 1.6 V and 2.0 -- 2.5 V produce samples with
stable phases of $x$ close to 0.33 and 0.25 respectively, which had been
difficult to produce due to the high current side reactions in previous
studies.\cite{Chou2005}  We also measured the open circuit
potential (OCP) after each equilibrium phase is reached as shown in
Fig. \ref{fig:deintercalation}. The OCP is slightly higher than
V$_{app}$ for x $>$ 0.5, but is significantly lower than V$_{app}$
for $x < 0.5$ due to the side reaction.  Since the OCP primarily
reflects the surface energy, the trend of OCP $> V_{app}$ for $x >
0.5$ suggests the bulk is equilibrating with a $V_{app}$ that is
lower than the surface energy.  The results shown in
Fig. \ref{fig:deintercalation} provide us with new insights into the
formation of phases below $x = 0.5$ using $V_{app}$ instead of OCP.
The plateaus in the $V_{app}-x$ plot indicate phase coexistence
regions with $dx/dV_{app} = 0$ around $\sim$ 0.5 ,1.3 and 2.0 V. The
corresponding stables phases are near $x = 0.50$, 0.33 and 0.25,
which are near the simple fractional fillings of 1/2, 1/3, and 1/4.
\begin{figure}
\includegraphics[width=3.3in]{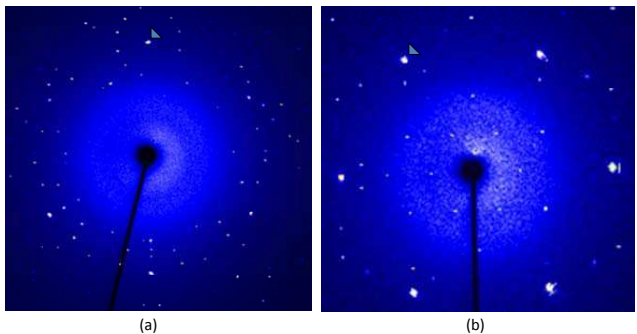}
\caption{\label{fig:Laue-084-025-2} (Color online) Representative Laue photographs
for (a) the starting crystal with $x = 0.84$ and (b) a
de-intercalated sample with the lowest Na content $x = 0.25$. The
six brightest spots have been identified to be \{100\} with $d =
2.83${\AA} indicated by the small triangle.  For $x = 0.84$, there
are 12 spots surrounding \{100\} in a ring-like pattern.  For $x =
0.25$, orthorhombic superlattice peaks can be identified in addition
to  \{100\}.}\vspace{-5mm}
\end{figure}
Single crystal quality is demonstrated by transmission Laue
photography as shown in Fig. \ref{fig:Laue-084-025-2} using
synchrotron X-rays ($\lambda$ = 0.4959 ${\AA}$), where $x = 0.84$ is
the original as-grown crystal and $x = 0.25$ is the most
de-intercalated sample.  A Na superstructure indicated by 12 peaks
forming a ring surrounding (100) is observed, just like that
reported in Ref.\cite{Roger2007a}.  An orthorhombic superstructure
is found in $x = 0.25$, which is different from that observed in $x =
0.5$ with electron diffraction.\cite{Zandbergen2004}.  Details of
the superstructure analysis will be presented separately.\cite{Chou}

In addition to the expected phases such as $x = 1/2$, 1/3 and 1/4,
stable phases of $x = 0.43$, 0.71 and 0.75 are also clearly visible
in Fig. \ref{fig:deintercalation}.  Although most early reports
suggest that $x = 2/3$ is a preferred equilibrated phase, it failed
to show up in the $V_{app}$ versus $x$ data.  This result is in
agreement with the prediction from Density Functional Theory (DFT)
calculation, which suggests $x = 2/3$ is not a ground state due to the lack of a commensurate Na ordering pattern.\cite{Zhang2005}  Our
data indicate that $x = 0.71$ is the most stable phase among 0.75,
0.71 and 0.67, as reflected on the significantly wider applied
voltage range of the $dx/dV_{app}=0$ plateau.  The stable phase of
$x = 0.71$ can be prepared with $V_{app}$ between -0.1 -- 0.1V,
while $x = 0.75$ is barely defined by a plateau of width less than
60 mV.  The $x = 0.75$ phase has been shown to be at the lower
boundary of the range ($0.75 \leq x \leq 0.85$) for compositions
that possess A-type AF ordering at low
temperature.\cite{Mendels2005, Sakurai2004b} In addition, $x = 0.75$ falls at the
exact boundary that minor $x$ fluctuations would separate Na(2) to
occupy either high symmetry 2c (2/3, 1/3, 1/4) or low symmetry 6h
(2x, x, 1/4) sites within the space group P6$_3$/mmc.\cite{Huang2004b}
Such behavior indicates that $x = 0.75$ is likely to be a metastable
transient phase rather than a truly equilibrated ground state. The $x =
0.71$ phase is very close to a simple fractional filling of 5/7 that
has been predicted to be a ground state from Density Function Theory
(DFT) model calculation;\cite{Zhang2005} however, there are many
other choices for a simple fraction that are near 0.71. Roger {\em
et al.} have proposed an ordering of multi-vacancy clusters within
the Na layer for $x > 0.7$, i.e., Na ions shift from the preferred
Na(2) site to the unfavorable Na(1) site that is directly on top of
the Co ions to further reduce stabilization energy.\cite{Roger2007a}

Another equilibrated phase below half filling is $x = 0.43$.
Although the $dx/dV_{app}=0$ plateau for $x = 0.43$ is less
pronounced than that for $x = 0.75$, the narrow plateau sits between
two distinctly different slopes as shown in
Fig. \ref{fig:deintercalation}.  Considering the dominant diffusion
species for x $>$ 0.5 are vacancy clusters, the diffusion species
for $x = 0.43$ would be Na ions.  The fact that 0.43 being close to
a simple fractional filling of 3/7 suggests that Na ions partially
fill a superlattice consisting of 7 Na ions per unit supercell.
Indeed Na trimer ordering of simple hexagonal unit \textbf{a}' =
$\sqrt{7}$\textbf{a} has recently been observed on the surface of an
in-situ cleaved Na$_{0.84}$CoO$_2$ crystal using scanning tunneling
microscopy, where half of the Na ions (x $\sim \frac{0.84}{2}$) are
supposed to occupy each cleaved surface pair equally.\cite{Pai} Whether
such 3/7 superlattice ordering exists within the bulk requires
further Laue transmission diffraction analysis.
\begin{figure}
\includegraphics[width=3.3in]{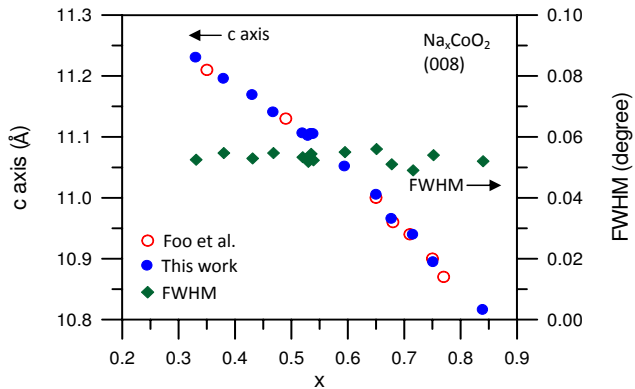}
\caption{\label{fig:c-axis} (Color online) The $c$-lattice parameter of
Na$_x$CoO$_2$ plotted against $x$. Crystal homogeneity is confirmed
by the FWHM in $2 \theta$ of the (008) Bragg peak.  The instrumental
resolution is 0.05 degrees.} \vspace{-5mm}
\end{figure}
\subsubsection{\label{sec:level2}Phase homogeneity}
One of the major concerns in the experimental study of Na$_x$CoO$_2$
compounds is the question of phase homogeneity.  When Na ions are
de-intercalated through topochemical or electrochemical routes, the
remaining Na  ions should rearrange to find the ground state for each
Na layer while maintaining the original CoO$_6$ stacking.  The
$c$-lattice parameters for our series of crystals were determined
with X-ray diffraction and are summarized in Fig. \ref{fig:c-axis}.
Since Na$_x$CoO$_2$ samples have highly oriented at (00L) planes
even after the crystal is pulverized, X-ray diffraction has been
performed on aligned crystals measuring the (00L) Bragg peaks. The
$c$-axis is known to strongly depend on Na content as reported by
several other groups.\cite{Foo2004, Chen2004, Prabhakaran2003} Our
$c$-axis lattice size is extracted from the (008) peak angle, and
our results are consistent with those reported
previously.\cite{Foo2004} We notice the $c$-lattice parameter
depends almost linearly on $x$ with a weak deflection near $x \sim
0.55$.  This behavior may suggest two regimes which follow the
empirical Vegard's rule for substitutional impurities. The $c$-axis
expansion as a function of lower $x$ has also been reflected on the
softening A$_{1g}$ mode along the $c$-direction.\cite{Lemmens2006}  As
shown in the inset of Fig. \ref{fig:c-axis}, phase homogeneity is
evidenced by a consistently narrow full width at half-maximum (FWHM)
for $2\theta$ of the (008) peak.  This width is comparable to our
instrumental resolution limit, which is determined using a Si
standard to be 0.05 degrees.
\begin{figure}
\includegraphics[width=3.3in]{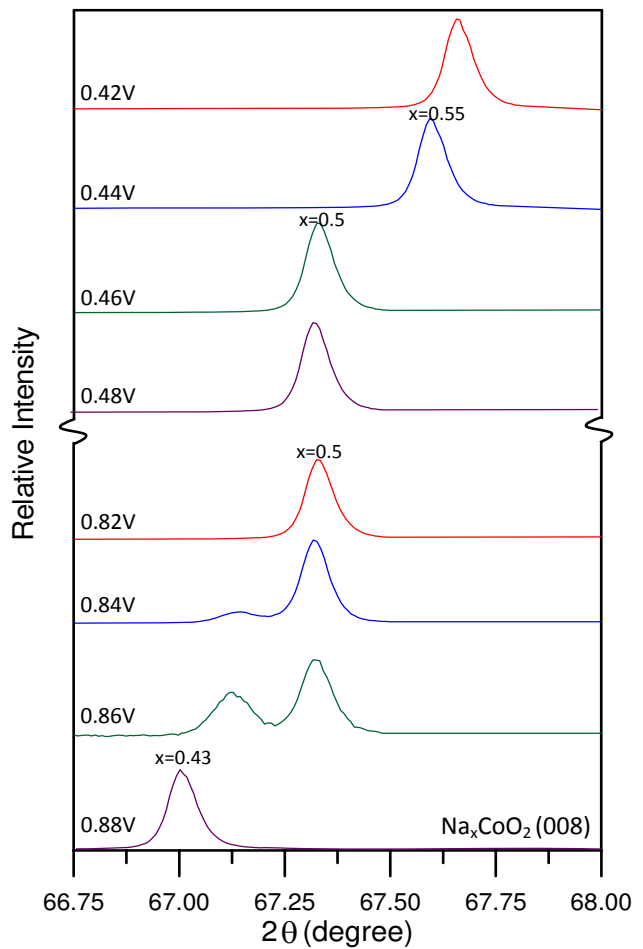}
\caption{\label{fig:2phase} (Color online) X-ray diffraction scans of the (008)
peak for different applied voltages near $x = 0.5$.  For $x \geq
0.5$, all of our crystals appear to be single phase.  For $x \leq
0.5$, two-phase coexistence is observed for the range between $x =
0.5$ and $x = 0.43$}. \vspace{-5mm}
\end{figure}
The possibility of mixed-phase crystals is ruled out within our
X-ray resolution limit for nearly all of the samples prepared by our
method, except for samples prepared using $V_{app}$ between 0.82 --
0.88 V as shown in Fig. \ref{fig:2phase}.  Two-phase behavior is
observed within the narrow voltage range between 0.82 -- 0.86 V,
which corresponds to Na levels slightly below $x=0.5$.  The sample
evolves from single phase of $x = 0.5$ to a single phase of $x =
0.43$ through mixed-phase intermediate states. In contrast, there is
no two-phase signature observed between single phase $x = 0.55$ and
$x = 0.5$. No other samples exhibited two-phase signatures
regardless of whether V$_{app}$ falls in the $dx/dV_{app}=0$ plateau
range.  This suggests that phases of solid-solution without specific
Na ordering do exist, at least in the time frame of days
during preparation and characterization.  Since the FWHM of (00L)
peaks provide us only with coherence information along the $c$-axis,
it is possible that in-plane ordering may yield potential
microscopic phase separated domains as commonly found in this
series.\cite{Zandbergen2004, Yang2004a}
\begin{figure}
\includegraphics[width=3.3in]{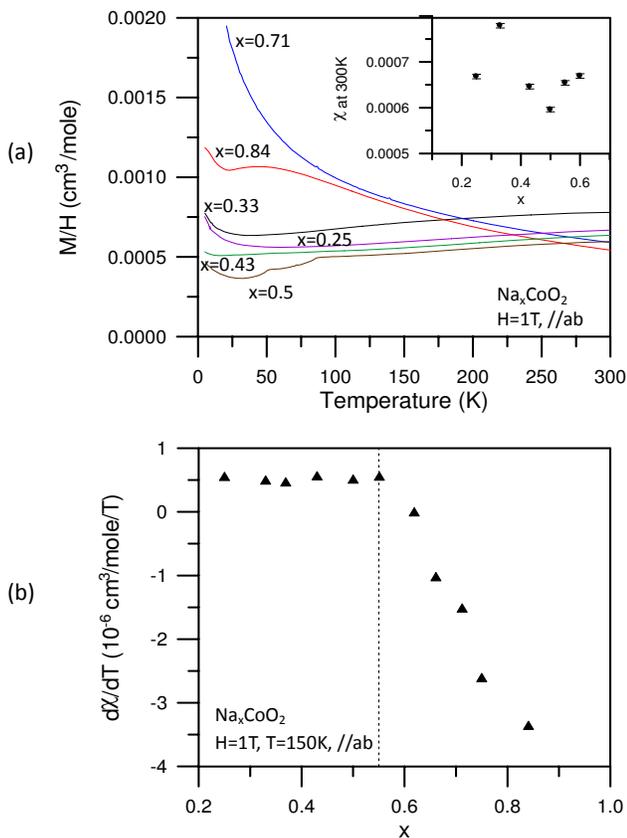}
\caption{\label{fig:SQUID-all-2} (Color online) (a) Magnetic susceptibility versus
temperature for crystals measured under applied field of 1 Tesla
along the $ab$-direction. The value of the  susceptibility at
$T=300$~K has been plotted in the inset, which has a maximum near $x
= 0.33$ and a minimum near $x = 0.50$. (b) The slope of the average
susceptibility versus $T$ at 150 K is plotted against $x$.  There
appears to be a crossover near $x = 0.55$.} \vspace{-5mm}
\end{figure}
\subsubsection{\label{sec:level2}Magnetic properties}
Additional evidence of phase purity is provided by magnetic
susceptibility measurements. Figure~\ref{fig:SQUID-all-2} displays
the complete susceptibility data measured under a magnetic field of
1 Tesla applied along the $ab$-direction. These results are
consistent with most of the studies reported so far: magnetic behavior
evolves from Curie-Weiss-like to Pauli-like as Na ions are
de-intercalated from 0.84 to 0.25, without any trace of CoO or
Co$_3$O$_4$ impurity.\cite{Chou2004a, Foo2004, Prabhakaran2003,
Yokoi2005} Assuming identical Van Vleck susceptibility contribution
for the Pauli paramagnetic metal region, the temperature independent
portion of the average susceptibility ($\chi$(300K)) should be
proportional directly to the density of states at the Fermi level.
The susceptibility at $T=300$ K $\chi$(300K) for $x \leq 0.6$ are
shown in the inset of Fig. \ref{fig:SQUID-all-2}, where $\chi$(300K)
for $x = 0.33$ appears as a local maximum and x = 0.5 sits at the
local minimum.  This trend agrees well with $\gamma$ values
extracted from specific heat measurement, where $\gamma$ is directly
proportional to the density of states at Fermi level.\cite{Jin2005}
The enhancement of density of state at Fermi level for $x = 0.33$
reflects its unique superconducting properties after
hydration.\cite{Takada2003} The local minimum of $\chi$(300K) for $x
= 0.5$ also indicates its proximity to the insulating state that
sets in below $T=51$~K.\cite{Foo2004}  The slope of the spin
susceptibility at 150K in Fig. \ref{fig:SQUID-all-2}a is an indicator
that separates Curie-Weiss (negative) and Pauli (positive)
behaviors.\cite{Yokoi2005}  The crossover point is close to $x =
0.55$ as shown in Fig. \ref{fig:SQUID-all-2}b, which is consistent with
the crossover points that have been observed in
Fig. \ref{fig:diffusion} of Na self diffusion and $c$-lattice
parameter as shown in Fig. \ref{fig:c-axis}.
\begin{figure}
\includegraphics[width=3.3in]{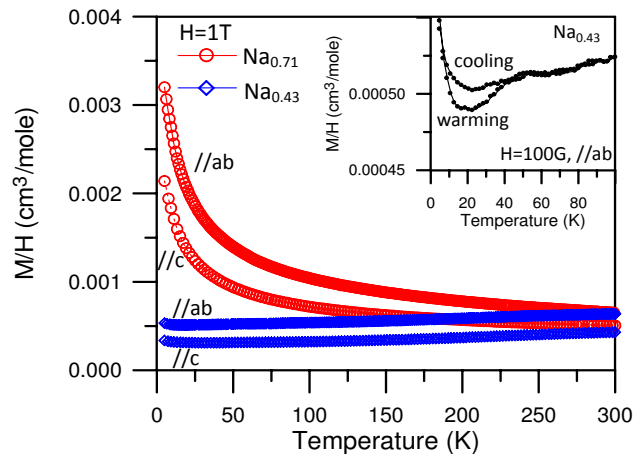}
\caption{\label{fig:SQUID-071-043} (Color online) Magnetic susceptibility versus
temperature for Na$_{0.71}$CoO$_2$ and Na$_{0.43}$CoO$_2$, with $H =
1$ Tesla for both $H//c$ and $H//ab$ directions. The inset shows a
low field measurement for Na$_{0.43}$CoO$_2$ using 100 Oe under
zero-field-cooled and field-cooled conditions.} \vspace{-5mm}
\end{figure}
In Fig. \ref{fig:SQUID-all-2}a, the in-place susceptibility
$\chi_{ab}$(T) for $x = 0.71$ and $x = 0.84$ are nearly identical
above $\sim$ 100K, but x = 0.71 does not show any magnetic ordering
down to 1.8~K. This behavior suggests $x = 0.71$ is the first
equilibrated phase below $x = 0.85$ that does not possess the well
known A-type AF ordering below 22-28 K.\cite{Bayrakci2005,
Mendels2005} Magnetic properties of the newly identified
Na$_{0.71}$CoO$_2$ and Na$_{0.43}$CoO$_2$ compositions are
summarized in Fig. \ref{fig:SQUID-071-043}. The Curie-Weiss-like low
temperature upturn for $x = 0.71$  agrees very well with the recent
$^{23}$Na NMR Knight shift data on samples with $x \sim
0.7$.\cite{Mukhamedshin2007} Powder average of $\chi_{ab}$ and
$\chi_c$ can be fitted to a simple relationship of
$\chi(T)=\chi_\circ+C/(T-\Theta$).\cite{Chou2004a}  Fitting over the
temperature range $T=50-250$~K yields $C = 0.078$ cm$^3$K/mole and
$\Theta$ $\sim$ -42K.  In a simple ionic picture, this corresponds
to about a 72\% concentration of Co$^{4+}$ ions that participate in
the weak antiferromagnetically correlated paramagnetic behavior.
Then the electrons on the remaining $\sim$28\% of the Co ions are
itinerant.  We speculate that a special Na ion ordering of $x =
0.71$ may form a large supercell that provides the underlying
potential for such  partial charge delocalization.

Na$_{0.43}$CoO$_2$ shows Pauli-like behavior just like all other
samples with $x \leq 0.55$. However, low field measurement revealed a
weak thermal hysteretic behavior below $\sim$40 K as shown in the
inset of Fig. \ref{fig:SQUID-071-043}.  The hysteresis onset near 40K
for $x = 0.43$ is very close to the metal-to-insulator transition
found in $x = 0.5$, which seems to suggest the inclusion of nearby
$x = 0.5$ phase.  In addition, there are traces of weak 51 K and 88 K
anomalies observable at the level of a few percent.  However, the
transition at $\sim$51 K for $x = 0.5$ does not show thermal
hysteresis, instead, the onset of hysteresis is found to set in below
the weak ferromagnetic phase $\sim$27K.\cite{Gasparovic2006}  Based
on the room temperature structural evidence shown in
Fig. \ref{fig:2phase}, we can produce samples with no $x = 0.5$ phase
inclusion within the $x =0.43$ matrix. We cannot rule out the
possibility of phase segregation below room temperature. The
observed thermal hysteresis of the susceptibility below $\sim$40 K
may be related to either weak ferromagnetism or spin glass behavior.
\begin{figure}
\includegraphics[width=3.3in]{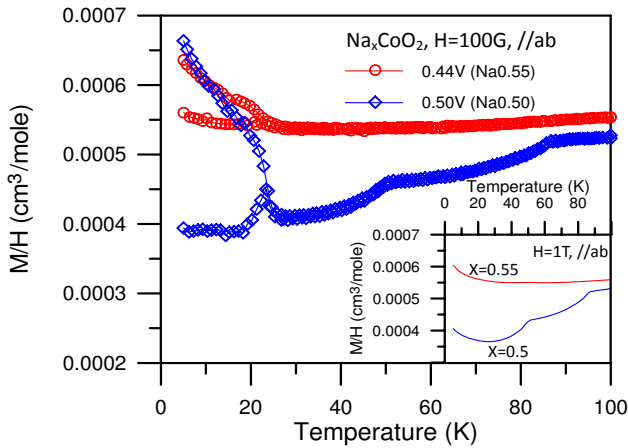}
\caption{\label{fig:SQUID-055} (Color online) Magnetic susceptibility versus
temperature for Na$_{0.55}$CoO$_2$ and Na$_{0.5}$CoO$_2$ prepared
using 0.44 V and 0.50 V for 24 hours each. The applied field is 100
Oe along the $ab$-direction, and measurements were taken under
zero-field-cooled and field-cooled conditions. The inset shows
similar measurements with a higher field of 1 Tesla.} \vspace{-5mm}
\end{figure}
For Na$_{0.5}$CoO$_2$, three phase transitions can be identified at
$\sim$27, 51 and 88 K in the magnetic susceptibility measured along the
$ab$-direction under low magnetic field, as shown in
Fig. \ref{fig:SQUID-055}. The transition at 88 K is due to a novel
antiferrmagnetic ordering transition, where the ordered spin
arrangement and direction has been deduced from polarized neutron
diffraction.\cite{Gasparovic2006} The transition at 51 K correlates
to a metal-to-insulator transition which may be related to nesting
of the Fermi surface.\cite{Bobroff2006} Most of the magnetic
measurement results of Na$_{0.5}$CoO$_2$ at high field before show
low temperature upturns below $\sim$30~K of various sizes, usually
attributed to trace amount of isolated paramagnetic
spins.\cite{Watanabe2006, Pedrini2005} Interestingly, we find the
transition at 27 K shows ferromagnetic character with thermal
hysteresis at low field as well as field hysteresis at 5~K (not
shown).  The significant thermal hysteresis found below $\sim$27K
for Na$_{0.5}$CoO$_2$ strongly implies its weak ferromagnetic
nature.  This transition below 27~K appears to further increase the
in-plane resistivity.\cite{Gasparovic2006}

The Na$_{0.55}$CoO$_2$ composition has been previously reported to
have a strong in-plane ferromagnetic ordering below $\sim$20K with
no magnetic anomalies found near 51 or 88 K.\cite{Wang2006} One
immediate concern on preparing pure Na$_x$CoO$_2$ crystal with $x =
1/2$ is how to separate the potentially mixed phases between
Na$_{0.55}$CoO$_2$ and Na$_{0.5}$CoO$_2$. A series of fully
equilibrated crystals have been prepared with different applied
voltage in small steps between 0.40 -- 0.50 V; the results of structure
evolution are shown in Fig. \ref{fig:2phase} and magnetic
susceptibilities are plotted together with that of $x = 0.5$ in
Fig. \ref{fig:SQUID-055}. High field susceptibility measurements show
a clear difference: nearly temperature independent behavior for
$x=0.55$ to the one with positive slope for $x=0.5$.  Low field
measurements indicates our Na$_{0.55}$CoO$_2$ crystal has a weak
ferromagnetic transition near 27 K without any 51~K or 88~K
anomalies, in contrast to Na$_{0.5}$CoO$_2$. This rules out the
possibility of a mixed phase within such a small concentration
range.  Na$_{0.55}$CoO$_2$ and Na$_{0.5}$CoO$_2$ should be
considered as two distinctly different phases, although the former
is closer to a critical transient state with an extremely narrow
$dx/dV_{app}$ plateau as indicated in Fig. \ref{fig:deintercalation}.
We note that the saturated moment and coercive field for
Na$_{0.55}$CoO$_2$ obtained from the electrochemical method is lower
than that prepared by the chemical extraction method using
iodine/acetonitrile solution.\cite{Wang2006}  In Na$_{0.5}$CoO$_2$,
the size of the magnetic susceptibility upturn below $\sim$27 K
appears to be highly sample dependent in the
literature.\cite{Bobroff2006,Gasparovic2006,Pedrini2005,Yokoi2005}
One major difference between chemical and electrochemical methods has
been compared on Li$_x$CoO$_2$.  Chemical de-intercalation method has a
significantly higher charging rate, and thus easier to access the
metastable states.\cite{Venkatraman2002}  We believe the chemical
method with a much higher charging rate may yield metastable phases
of Na$_{0.55}$CoO$_2$ of higher weak FM moment, and these may
eventually fall into a more equilibrated phase similar to that
obtained using the electrochemical route.
\subsubsection{\label{sec:level2}Transport properties}
\begin{figure}
\includegraphics[width=3.3in]{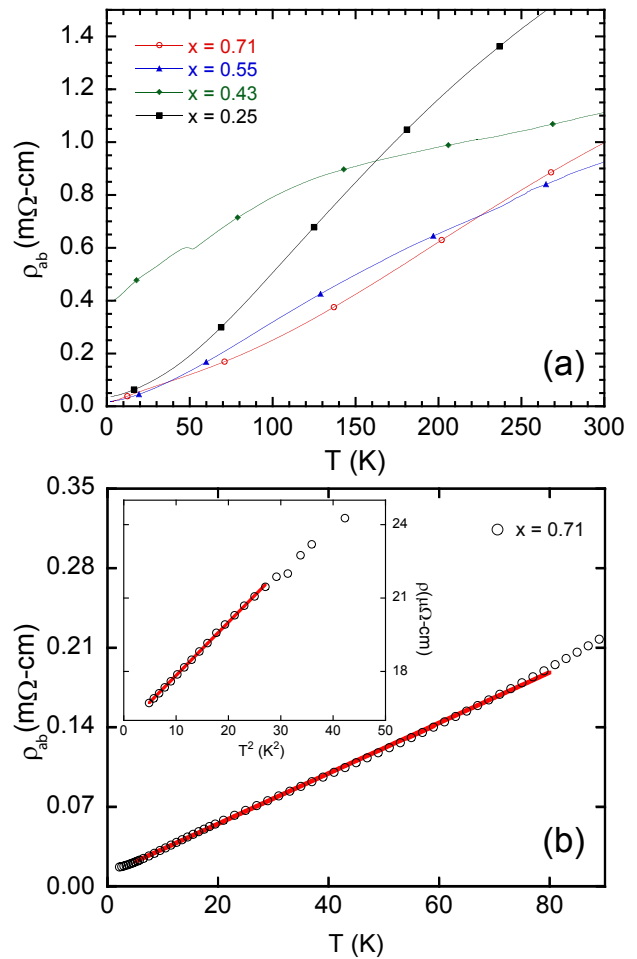}
\caption{\label{fig:res} (Color online) (a) Temperature dependence of the in-plane
resistivity for several of the Na$_{x}$CoO$_2$ crystals in the
series. (b) Low temperature resistivity of the composition with $x =
0.71$.  Two regimes are clearly seen: a $T^2$ temperature dependence
below 5~K (inset) and a $T$-linear temperature dependence between
$\sim 5$~K and 80 ~K.} \vspace{-5mm}
\end{figure}
The transport properties of four of our newly synthesized
compositions are summarized in Fig.~\ref{fig:res}.  As shown in
Fig.~\ref{fig:res}(a), the compositions with $x=0.71,0.55,$ and
$0.25$ exhibit metallic conductivity, and are consistent with the trend
suggested by the phase diagram previously proposed.\cite{Foo2004}
The sample with $x = 0.43$, does not behave like a good metal and
appears to have a large residual resistivity. This particular sample
may contain coexisting regions of different phases, such as
$x=0.5$, which are more stable at the local scale. Indeed, after
annealing at $T_a = 400$K for several hours (not shown), this sample
displays a transition to an insulating regime below $T \approx 50$K,
which is a particular feature of the $x=0.5$
composition\cite{Gasparovic2006}.
The resistivity of the $x = 0.71$ sample at lower temperatures is
shown in detail in Fig.~\ref{fig:res}(b).  A roughly $T$-linear
temperature dependence is observed between $\sim 5$~K and 80~K,
consistent with previous measurements.\cite{Foo2004}  Fermi liquid
behavior is recovered at low temperatures where electron-electron
scattering dominates, as noted previously.\cite{Li2004a}  Here,
below $\sim 4$K the data may be fit by the function of $\rho (T)=
\rho_\circ + AT^2$. Our fit (plotted as the solid line) yields
values of $\rho = 15.61~\mu\Omega-cm$ and $A = 0.2226~
\mu\Omega-cm~K^{-2}$.  The residual resistivity $\rho_\circ$ may
arise from scattering due to domain boundaries and/or perturbations
to the periodic potential felt by electrons within the $ab$ plane,
such as defects in the Na arrangement. The observed value of
$\rho_\circ$ is about three times smaller than previous
measurements\cite{Li2004a} and attests to the high homogeneity of
the current Na-ordered $x=0.71$ sample. The value of $A$, while
somewhat smaller than previous measurements, is still relatively
large and indicates an large Kadowaki-Woods ratio, $k_{KW} = A /
\gamma^2$, assuming $\gamma$ is similar to previous
measurements.\cite{Li2004a}
\\
\section{\label{sec:level1}Summary\protect\\}
In summary, we have demonstrated that a non-aqueous electrochemical
de-intercalation process can be used to prepare high quality single
crystal samples of $\gamma$-Na$_x$CoO$_2$ with well-defined Na
superlattices.  The sodium ion self diffusion mechanism has been
demonstrated to be well fit by a 1D diffusion model on an aged
single crystal. The dominant diffusion species above $x_c \sim 0.53$
is consistent with Na vacancies, and the diffusion coefficient is
higher for $x>x_c$, compared to $x<x_c$.  The existence of stable
phases are reflected in the $dx/dV_{app}=0$ plateaus, among which $x
= 1/2$, 1/3 and 1/4 are the most stable.  The data suggest that $x
\sim 0.75$, 0.55 and 0.43 are somewhat less stable.  On the other
hand, $x\sim 0.71$ is a very stable phase with a sizeable
$dx/dV_{app}=0$ plateau width.  This stable $x = 0.71$ phase is
intriguing for its potential Na ordering but without any magnetic
ordering down to 1.8K.  The absence of A-type AF ordering for $x =
0.71$ and the presence of weak FM in $x = 0.5$ and 0.55 may be
related to subtle interactions between the Co ions and the specific
potential caused by Na ion ordering, as it is known that Na
superstructures can affect the Fermi surface geometry. Clearly more
work is required to investigate the effects of Na order across the
Na$_x$CoO$_2$ phase diagram.  Our progress with this electrochemical
technique opens the door for further studies of the physical
properties of the stable Na ordered phases in single crystal
samples.
\\
\noindent$^\dagger$corresponding author: fcchou@ntu.edu.tw
\
\begin{acknowledgments}
FCC and GJS acknowledge B. X. Xie's help on fitting of diffusion
equation.  We acknowledge the helpful discussions with M. W. Chu, Larry
Pai, and Patrick Lee.  This work was support by the National Science
Council of Taiwan under project number NSC-95-2112-M-002.  The work
at MIT was supported by the U.S. Department of Energy under Grant
No. DE-FG02-04ER46134.
\end{acknowledgments}

\end{document}